\documentclass[aps,showpacs,floatfix,twocolumn,byrevtex,superscriptaddress,longbibliography,prb]{revtex4-1}
\usepackage{amsmath}
\usepackage{amssymb}
\usepackage{amstext}
\usepackage{amsopn}
\usepackage{amsfonts}
\usepackage{amsxtra}
\usepackage{amsmath,amsfonts,amssymb}
\usepackage{xcolor}
\usepackage{lipsum}
\usepackage[english]{babel}
\usepackage{graphicx}
\usepackage{float}
\usepackage{bm}
\usepackage{multirow}
\usepackage{dcolumn}
\usepackage[colorlinks]{hyperref}
\usepackage{color,soul}
\usepackage{url}
\usepackage{breakurl}
\usepackage{bm}

\newcommand{\h}{\hbar}

\newcommand{\ee}{\varepsilon}
\newcommand{\kk}{\mathbf{k}}

\newcommand{\s}{\sigma}

\begin{document}

\title{Optical conductivity of the type-II Weyl semimetal TaIrTe$_4$}

\author{F. Le Mardel\'e}
\email[]{florian.lemardele@unifr.ch} 

\author{D. Santos-Cottin}
\affiliation{Department of Physics, University of Fribourg, 1700 Fribourg, Switzerland}

\author{E. Martino}
\affiliation{Department of Physics, University of Fribourg, 1700 Fribourg, Switzerland}
\affiliation{IPHYS, EPFL, CH-1015 Lausanne, Switzerland}

\author{K. Semeniuk}
\affiliation{IPHYS, EPFL, CH-1015 Lausanne, Switzerland}

\author{S. Ben David}
\affiliation{Department of Physics, University of Fribourg, 1700 Fribourg, Switzerland}

\author{F. Orbani\'c}
\author{M. Novak}
\affiliation{Department of Physics, Faculty of Science, University of Zagreb, Bijeni\v{c}ka 32, HR-10000 Zagreb, Croatia}

\author{Z.~Rukelj }
\affiliation{Department of Physics, University of Fribourg, 1700 Fribourg, Switzerland}
\affiliation{Department of Physics, Faculty of Science, University of Zagreb, Bijeni\v{c}ka 32, HR-10000 Zagreb, Croatia}

\author{C.~C.~Homes}
\affiliation{Condensed Matter Physics and Materials Science Division, Brookhaven National Laboratory, Upton,
   New York 11973, USA}

\author{Ana  Akrap}
\email[]{ana.akrap@unifr.ch}
\affiliation{Department of Physics, University of Fribourg, 1700 Fribourg, Switzerland}

\date{\today}
\begin{abstract}
TaIrTe$_4$ is an example of a candidate Weyl type-II semimetal with a minimal possible number of Weyl nodes. Four nodes are reported to exist a single plane in $k$-space. The existence of a conical dispersion linked to Weyl nodes has yet to be shown experimentally. Here we use optical spectroscopy as a probe of the band structure on a low-energy scale.
Studying optical conductivity allows us to probe intraband and interband transitions with zero momentum. In TaIrTe$_4$, we observe a narrow Drude contribution and an interband conductivity that may be consistent with a tilted linear band dispersion up to 40~meV. The interband conductivity allows us to establish the effective parameters of the conical dispersion; effective velocity $v=1.1\cdot 10^{4}$~m/s and tilt $\gamma=0.37$.
The transport data, Seebeck and Hall coefficients, are qualitatively consistent with conical features in the band structure. Quantitative disagreement may be linked to the multiband nature of TaIrTe$_4$.
\end{abstract}
\maketitle

%%%%%%%%%%%%%%%%%%%%%%%%%%%%%%%%%%%%%%%%%%%%%%%%%%%%%%%%%%%%%%%%%%%%%%%%%%%%%%%
%
% Introduction
%\section{Introduction}

Weyl semimetals are currently intensely studied both by theoretical and experimental methods. Their effective description at low energies is given by the Weyl equation, which is a variant of Dirac equation for particles with zero rest mass. This effective description results in special properties of the wave functions, leading to the appearance of pairs of Berry monopoles. The effective Weyl description also leads to a linear dispersion of energy, usually constrained to a low energy range.

The first realization of Weyl nodes in a solid came in the TaAs family. However, the band structure in that family is far from simple, and the number of nodes is very high: 24.
In the past few years, efforts have been made to pinpoint a simpler realization of Weyl semimetal. Theoretically, the minimal allowed number of Weyl nodes is 4, for Weyl semimetals with broken inversion symmetry. These Weyl nodes are all located on a single plane in $k$-space. 
Arguably, TaIrTe$_4$ is an example of the theoretically predicted minimal Weyl semimetal, possessing only 4 Weyl nodes.\cite{Koepernik2016, Zhou2018} While {\em ab initio} calculations predicted the existence of Weyl nodes in TaIrTe$_4$, surface techniques reported signatures of Fermi arcs,\cite{Belopolski2017} despite the challenge of Weyl cones residing above the Fermi level. 
A number of exotic properties have been claimed to arise in TaIrTe$_4$, among them are topological edge states,\cite{Dong2019}
unconventional surface superconductivity,\cite{Xing2019} bulk superconductivity under pressure,\cite{Cai2019} large Fermi arcs reaching $1/3$ of the Brillouin zone dimension,\cite{Belopolski2017} stretchable Weyl points and line nodes,\cite{Zhou2018} and circular photogalvanic effect linked to the Berry curvature.\cite{Ma2019}

Unfortunately, many of these results hinge on the {\em ab initio} calculations, which become fairly unreliable at low energies on a scale of tens of milli-electron volts. One cannot trust the density functional theory (DFT) calculations to tell correctly what is the energy scale of the linear dispersion, which would be linked to the Weyl cones. Experimentally, it is difficult to expose those experimental features which are directly linked to the Weyl dispersion, particularly in the bulk properties of the candidate material. 
There are still only few explicit confirmations of the conical features in 3D materials such as TaIrTe$_4$. 
This is why, for TaIrTe$_4$, a number of questions remain unclear: Can one detect the conical dispersion and what is its scale? Where is it located with respect to the Fermi level? How much are these energy-band cones tilted?

In this work we determine the optical conductivity of TaIrTe$_4$. We focus in particular on its low-energy range, in order to identify the possible signatures of tilted Weyl dispersion, expected to be characteristic of this material.
While one inherently cannot say anything about the Weyl character of the bands from probing the simple optical response, one can reach conclusions about the energy dispersion and its possible linearity. 
Weyl cones in TaIrTe$_4$ are of type II, meaning that they are predicted to have a certain tilt. This tilt drastically changes properties, for example it removes the electron-hole symmetry. As a result, the optical response of a tilted cone will be different than of a non-tilted cone. The tilted conical dispersion may be simply described by a $2\times 2$ Hamiltonian matrix. From such a model, one can analytically derive various physical quantities. Specifically, here we derive the Seebeck coefficient or thermoelectric power, Hall coefficient, and the optical conductivity, in the tilted and non-tilted direction.
We show, using our optical spectrum calculation for a tilted 3D conical dispersion, that even in an apparently simple situation, there is some ambiguity in the parameters that describe this dispersion.
Results of our analysis are that the fingerprints of Weyl dispersion may exist on a scale of up to 40 meV.

%
% Figure 1
%
\begin{figure*}[!ht]
\includegraphics[width=\linewidth]{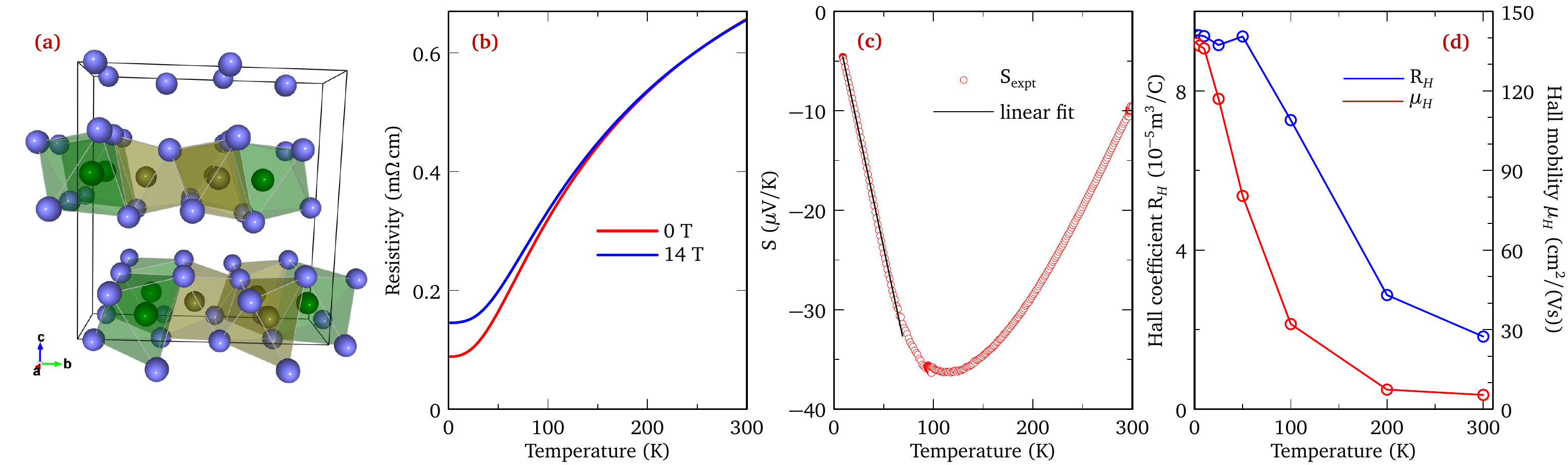}
\caption{(a) The unit cell of TaIrTe$_4$.  Purple balls are tellurium atoms. The green and brown octahedra contain, respectively, tantalum and iridium atoms. The metal-metal zigzag chains run along the $a$ axis.
(b) Resistivity as a function of temperature, in zero magnetic field and in 14~T. Magnetic field is applied along the $c$ axis.
(c) Thermoelectric power or Seebeck coefficient as a function of temperature. Black line shows a linear fit below 60~K, whose slope may be used to extract the Fermi level position. 
(d) Hall coefficient and Hall mobility as a function of temperature.
}
\label{fig1}
\end{figure*}
%

%
%Experimental and theoretical methods
%
Single crystals of TaIrTe$_4$ have been grown by a self-flux method. High purity starting elements in the ratio  Ta:Ir:Te=1:1:25 were sealed in a quartz tube under the vacuum better than 10$^{-5}$ mbar. The mixture was heated to $950^\circ$C (keeping it below of the boiling point of Te), allowing the flux to become homogeneous and slowly cooled for several days to $500^\circ$C. By centrifugation of the mixture shiny, easy to cleave,  needle-like single crystals of TaIrTe$_4$ with  the longitudinal dimension (along the b-axis) of several millimeters were obtained. Structure and composition of the crystals were confirmed using the X-ray diffraction and energy-dispersive X-ray spectroscopy. 
Electrical resistivity and Hall coefficient were measured in a Physical Property Measurement System from
Quantum Design as a function of temperature. The sample was measured using a four-probe
technique in a bar configuration in the $ab$-plane.
Thermoelectric power was measured using a home made setup.
The optical reflectivity was determined at a near-normal angle of incidence with light
polarized in the $ab$-plane for photon energies ranging between 2~meV and  2.7~eV
(16 and $22\,000$~cm$^{-1}$), at temperatures from 5 to 300~K.  The single crystal was
mounted on the cold finger of a He flow cryostat and absolute reflectivity was determined
using the \textit{in-situ} coating technique \cite{Homes_1993}.
The complex optical conductivity was obtained using a Kramers-Kronig
transformation from the reflectivity measurements. At low frequencies, we used a Hagen-Rubens
extrapolation. For the high frequencies, we completed the reflectivity data using the calculated
atomic X-ray scattering cross sections \cite{Tanner_2015} from 10 to 60~eV followed by a
$1/\omega^{4}$ dependence.

%
%\section{Results and discussion}
The crystal structure of TaIrTe$_4$ is shown in Fig.~\ref{fig1}a. Its orthorhombic structure is derived from the structure of WTe$_2$, with a doubling of the $b$ axis.\cite{Mar1992} TaIrTe$_4$ has the crystallographic space group $Pmn2_1$, with four formula units per unit cell. 
%\subsection{Transport results}
%

Figure~\ref{fig1}(b--d) shows the resistivity $\rho$, Seebeck coefficient $S$, and Hall coefficient $R_H$, each as a function of temperature. 
The resistivity in Fig.~\ref{fig1}b is metallic, with a residual resistivity ratio $\mathrm{RRR}=\rho(300\, {\mathrm K})/\rho(0\, {\mathrm K})= 7.4$. This value is much smaller than in WTe$_2$, but comparable to the results in literature.\cite{Khim2016, Xing2019}
The resistivity is somewhat enhanced in magnetic field, mostly at low temperature. The effect is much smaller than in WTe$_2$,\cite{Ali2014} ZrTe$_5$,\cite{Shahi2018} or in MoTe$_2$.\cite{Chen2016} Although the three systems are similar to TaIrTe$_4$ to a certain extent, an important difference is that this compound is not nearly as close to a perfect compensation between electron and hole pockets.

The Seebeck coefficient, Fig.~\ref{fig1}c, is negative in the entire temperature range. $S(T)$ has a specific shape given by a linear dependence at low temperatures, which is followed by a broad minimum around 120~K, and an increase towards room temperature. 
The linear $S(T)$ can be understood, in a first approximation, through the Mott expression which describes the diffusion of electrons in a parabolic energy band. The Mott formula relates Seebeck coefficient to the Fermi level, 
$S_{{\mathrm Mott}}(T)=k_B^2 T/(e \varepsilon_F)$. 
Through this formula, the experimental $S(T)$ would indicate that the Fermi level is $\varepsilon_F \sim 100$~meV. 
%
% Figure 2
%
\begin{figure*}[!ht]
\includegraphics[width=\linewidth]{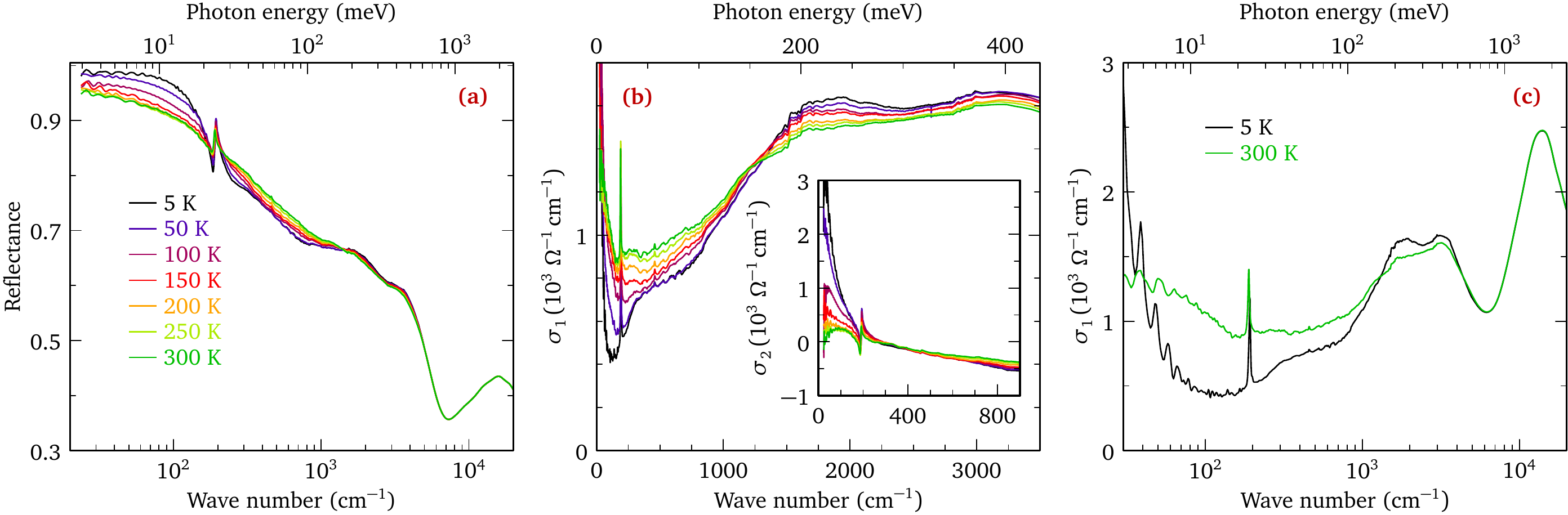}
\caption{(a) Reflectance measured at near-normal incidence for light polarized in the $ab$ plane, at several different temperatures.
(b) The real part of the optical conductivity, $\sigma_1$, as a function of incident photon energy, in the far infrared and mid infrared range (up to 400~meV), shown for the same set of temperatures as in (a). Inset shows the imaginary part of the optical conductivity, $\sigma_2$, below 900 cm$^{-1}$. 
(c) Low- and high-temperature $\sigma_1$ as a function of incident photon energy, shown in the entire experimental range, on a logarithmic scale.
}
\label{fig3}
\end{figure*}

Finally, the Hall coefficient $R_\mathrm{H}$, shown in Fig.~\ref{fig1}d, has the opposite sign from the Seebeck coefficient, in agreement with the literature.\cite{Khim2016} Its temperature dependence suggests an increase in the carrier density as the temperature increases.
It is known that TaIrTe$_4$ has a complex Fermi surface, composed of two electron and two nested hole pockets. \cite{Koepernik2016, Khim2016} A multiband situation means one cannot talk about electron- or hole-dominant conduction.
Fig.~\ref{fig1}d also shows the Hall mobility, $\mu_\mathrm{H} = \lim_{B \to 0} R_{xy}/(R_{xx} B)$. The value of $\mu_\mathrm{H}$ is low; even at 1.8~K, it remains below 140~cm$^{2}/$(Vs). 

Overall, the transport coefficients paint a picture of TaIrTe$_4$ as a poor metal, characterized by fairly high values of resistivity, small RRR, high values of Seebeck coefficient, but with a metallic temperature dependence. 
The quantum oscillation measurements, shown in Supplementary Materials,\cite{SM} confirm that  the Fermi surface is made out of several pockets.\cite{Khim2016} Through the transport measurements, we are able to access the intraband excitations, described by the Drude response. However, it is not possible based on transport alone to find the scattering rate, nor to identify if it is the carrier density, or the scattering rate, that causes most of the temperature dependence of the resistivity.

To address these questions, we resort to optical spectroscopy. The scattering rate and the carrier density are separated in the optical conductivity, as the strength and width of the Drude peak. The comparative advantage with respect to the transport measurements is not only the ability to disentangle different parameters of the Drude response, but also capability to access the excitations between different energy bands. 
These interband excitations directly probe the energy band dispersion through its joint density of states (JDOS).\cite{Martino2019}

%\subsection{Optical spectroscopy}

The reflectance of TaIrTe$_4$ at seven different temperatures is shown in Fig.~\ref{fig3}a. The response is decidedly metallic, with $R \rightarrow 1$ in the low-energy limit,  $\omega \rightarrow 0$. The temperature dependence is particularly strong in the far infrared range, and similar to what was observed in WTe$_2$ or MoTe$_2$.\cite{Homes2015,Santos-Cottin2020} A single strong, sharp phonon mode is detected at 24~meV (200 cm$^{-1}$).
Through Kramers-Kronig relations, one can determine the complex optical conductivity $\sigma = \sigma_1 + i \sigma_2$. The real part of this response function, $\sigma_1$, leads to absorption of radiation by the material. This component is of particular interest, because it is directly proportional to the JDOS.
On the contrary,  the imaginary component $\sigma_2$  does not result in absorption, but only leads to a phase lag.

The real part of the optical conductivity, $\sigma_1(\omega)$, is shown in Fig.~\ref{fig3}b for the low photon energies, and in Fig.~\ref{fig3}c for the broad energy range. Inset of Fig.~\ref{fig3}b shows the imaginary component, $\sigma_2(\omega)$.
A narrow Drude component develops below 10 meV, associated with the free carriers.
The Drude component is broader and therefore more evident in $\sigma_2(\omega)$. For example, at 300~K, a broad peak in $\sigma_2$ is centered around 100~cm$^{-1}$, the Drude scattering rate at that temperature. 

In $\sigma_1$, low-energy interband excitations dominate the optical response between 10 and 200~meV.
These excitations leave a characteristic signature which is intimately related to the JDOS, and may leave imprints of conical dispersion around the Weyl nodes.
Nearly linear dependence of $\sigma_1(\omega)$ extrapolates into a finite value at $\omega=0$. We will show later that such a finite offset is indeed expected for a tilted conical dispersion. In Fig.~\ref{fig3}c, the high energy excitations are also visible, centered at 0.2, 0.4 and 1.8~eV. Due to the complex band structure, it is not possible to assign these excitations to specific features of the band structure.\cite{Koepernik2016,Belopolski2017}

\begin{figure*}[!tb]
\includegraphics[width=\linewidth]{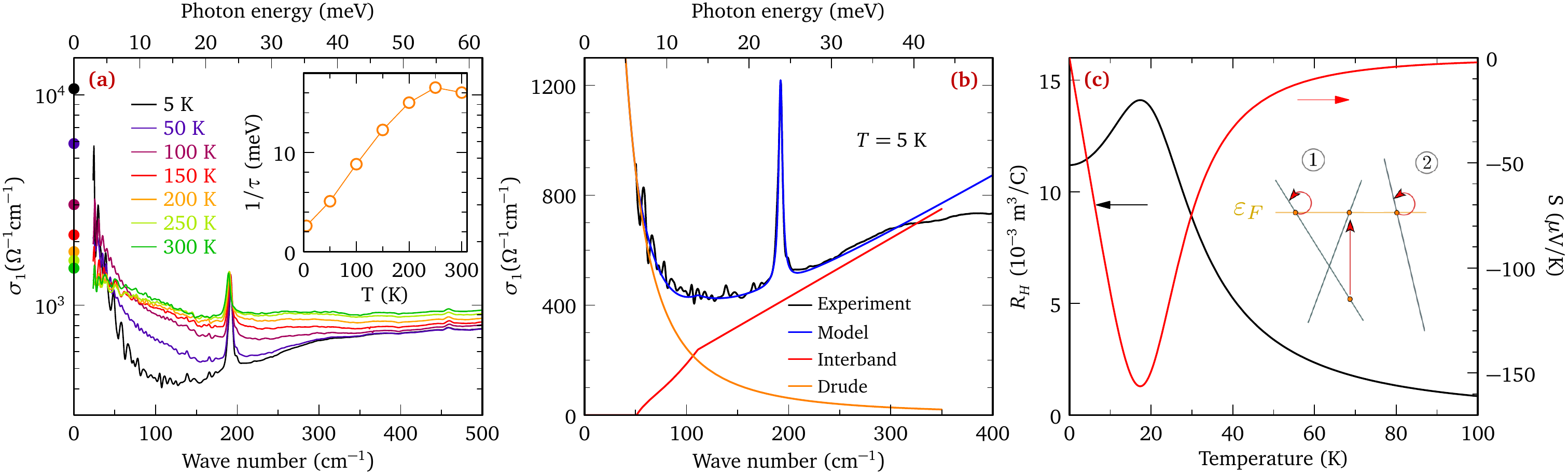}
\caption{(a) Optical conductivity in the far infrared range, showing the Drude component. The dots represent the values of $\sigma_{dc}$ determined by the measurement of the temperature dependent resistivity.
Inset shows the temperature dependence of the Drude scattering rate, $1/\tau$. It is extracted from $\sigma_1 (\omega)$ using a Drude-Lorentz model.
(b) Decomposition of $\sigma_1 (\omega)$ at 5~K into the Drude term, and an interband term. The model curve (blue) includes the Drude contribution, interband contribution from a tilted conical dispersion, and a phonon contribution.
(c) Theoretically calculated Hall coefficient $R_{\mathrm{H}}$, and Seebeck coefficient $S$ within the tilted conical model.
Inset in (c) shows a sketch of a tilted linearly dispersing band (1) and a trivial band (2); both bands cross the Fermi level. In the tilted cone, the low-energy interband transitions (vertical arrow) contribute to the optical conductivity. There is no contribution to the interband conductivity from the sketched trivial band. Both bands, however, contribute to the intraband conductivity, which is related to the measured $S$ and $R_H$.
}
\label{fig4}
\end{figure*}

Intraband excitations can be described by the Drude component, which may be characterized by a plasma frequency $\omega_{p,{\mathrm{Drude}}}$ and a scattering rate $1/\tau$.
The narrow Drude component in TaIrTe$_4$ is well distinguished in our measurements. This is seen in Fig.~\ref{fig4}a, which shows $\sigma_1(\omega)$ in the far infrared region. The dynamical conductivity is consistent with the $\sigma_{dc}$ values  extracted from the resistivity measurement in Fig.~\ref{fig1}b. The Drude component strongly narrows at low temperatures, similar to what is generally seen in similar compounds.\cite{Homes2015,Santos-Cottin2020,Martino2019} 
The scattering rate is shown in the inset of Fig.~\ref{fig4}c as a function of temperature. Expectedly, it increases with temperature.
 %The Drude weight is related to the effective carrier density, $\omega_{p,{\mathrm{Drude}}}^2 \propto n_{eff}/m_e$. While the scattering rate expectedly increases with temperature, the Drude weight shows a small increase ($\sim20 \%$). This increase in the carrier density is in agreement with the temperature dependence of $R_H$ in Fig.~\ref{fig1}c.
%
At low temperatures, the Drude plasma frequency is $\omega_{p,{\mathrm{Drude}}}=0.1$~eV, and the Drude scattering rate is $2.6$~meV. For comparison, similarly small scattering rates $\hbar/\tau \lesssim 1$~meV were found in related compounds \cite{Homes2015,Santos-Cottin2020} MoTe$_2$ and WTe$_2$, as well as in other candidate topological semimetals.\cite{Akrap2016,Crassee2018,Xu2016,Martino2019}
Such a small value is characteristic of a compound with a small Fermi surface, resulting in a small $k$-space available for intraband scattering. 

%
% Figure 3
%
%%Very generally, we can link the real part of the optical conductivity to the joint density of states, $\sigma_1 \propto \mathrm{jDOS}/\omega$. One can define jDOS as the density of states that corresponds to the energy difference between two bands, \hl{(formula)}. It may be evaluated knowing the energy dispersion. 

We distinguish three parts in $\sigma_1$ at low energies: a narrow Drude contribution, a sharp phonon contribution, and an interband part. Separating the Drude and phonon contributions allows to access the interband optical response: $\sigma_{1,\mathrm{interband}} = \sigma_1 - \sigma_{1,\mathrm{Drude}}- \sigma_{1,\mathrm{phonon}}$
The premise is that at low temperatures this interband response contains a signature of tilted conical dispersion linked to Weyl nodes. 
We perform the low energy optical analysis under certain assumptions. First, there are multiple bands at Fermi level, as inferred by DFT, but there are no interband transitions below 50~meV (400~cm$^-1$) except from those arising within the tilted cone. Second, the sum of all band contributions to the conductivity is contained in the $\sigma_{dc}$ and $\tau$ of the single Drude Lorentzian. 
Third, the footprint of any linear-dispersion-like signature is found in the data interval from 15 to 35~meV, where we see a linear-like conductivity (see Fig.~\ref{fig4}b). 

To identify the response from a possible Weyl dispersion, a numerical calculation can be done, and the parameters used to describe the Weyl cones can be adapted to best describe the experimental optical conductivity.
The Hamiltonian for a tilted Weyl cone is: 
\begin{equation} \label{ham1}
\hat{H}_0 = \h wk_z \mathcal{I} \ + \h v \kk \cdot \boldsymbol{\sigma},
\end{equation}
where $\s_{x,y,z}$ are Pauli matrices, $\mathcal{I}$ the identity operator, $v$ Dirac velocity and $w$ tilt velocity.
Diagonalizing the Hamiltonian gives its eigenvalues for conduction $c$ and valence $v$ bands,
%\begin{equation}\label{ham2}
 $\ee^{c,v}_{\kk} = \h w k_z \pm \h v |\kk|$.
%\end{equation}
Using these eigenvalues one can numerically calculate the optical conductivity.\cite{SM} We take into account four Weyl nodes, as given by DFT. The result is given in Fig.~\ref{fig4}c in red and blue lines, with the following parameters: Fermi level, with respect to the tip of the cone, $\varepsilon_F = 4.3$~meV, tilt of the cone $\gamma = w/v = 0.37$, and Fermi velocity $v=1.1 \cdot 10^4$~m/s.
The parameter $\gamma < 1$ that we obtain describes an under-tilted Weyl cone.
Although favoured by the DFT, over-tilted Weyl scenario ($\gamma > 1$) would give a significantly different shape of the interband part, incompatible with experimental data.

The optical conductivity is consistent with the calculated interband conductivity within a tilted conical dispersion. This cannot be taken as a smoking gun of Weyl dispersion, as we cannot exclude other dispersions which would result in a similar interband conductivity. However, if the band structure indeed contains four Weyl cones at low energies, then our result gives their energy scale which is less than 40~meV.

The same effective Hamiltonian model can be used to calculate the Seebeck coefficient and the Hall constant. These are shown in Fig.~\ref{fig4}c. Both are calculated  assuming that $\varepsilon_F = 4.3$~meV. 
While there is some agreement in the temperature dependence for both coefficients, quantitatively the model overestimates both $S$ and $R_{H}$.
%On a more absolute scale, we do not know where the cones are with respect to the zero of the energy band diagram. This can be approached using our transport data, namely the Seebeck coefficient. 

Although generally a complex quantity,\cite{SM} the Seebeck coefficient is simplified dramatically for a system with linear bands, when the effective Fermi velocity is constant.
In such a case, $S =(1/e) \, \partial \mu(T)/\partial T$, where $\mu(T)$ is the chemical potential.\cite{SM} 
This generically gives $S(T) \propto T$ at low temperatures, after which $S(T)$ reaches an extremum value, and then it decreases in absolute value. 
One interesting consequence of the above simple expression is that the minimum of $S(T)$ depends linearly on the Fermi energy, measured from the tip of the cone (or the Weyl node). In the case of TeIrTe$_4$, the experimental thermopower (Fig.~\ref{fig1}c) only qualitatively resembles the calculated $S(T)$, shown in Fig.~\ref{fig4}c. The experimental minimum of $S(T)$ is observed at 100~K, and the absolute value is about 5 times smaller than in the calculation. 
%This discrepancy may be in part caused by the multiband nature of TaIrTe$_4$.
%it is tempting to associate this minimum to the energy location of the Weyl node with respect to the Fermi level, which would give $\varepsilon_F \sim 100$~meV. In contrast, the optical data tells us that the Fermi level is about 12~meV away from the tip of the cone. 
TaIrTe$_4$ is a multiband system. While its optical conductivity may well be dominated by the interband transitions arising within the tilted cones, the intraband channel will contain contributions from all the bands that cross the chemical potential, as shown in the inset of Fig.~\ref{fig4}c. All these intraband excitations are reflected in the transport coefficients. Specifically, we see that $S(T)$ cannot be described by a simple linear-dispersion term shown in Fig.~\ref{fig4}c, and multiband contributions lead to a smaller experimental thermopower. Similarly, $R_H$ will be lower from what our model gives, if other bands at the Fermi level are considered, bringing it in closer agreement to the experiment.

%
%\section{Conclusions}
%
In conclusion, we have determined the optical conductivity of a candidate  type-II Weyl semimetal, TaIrTe$_4$. 
Through a combined use of detailed infrared spectroscopy and effective modelling, we show
that  the low-energy dynamical conductivity is broadly consistent with a tilted conical dispersion. This dispersion may explain the interband response at low energies, below 40~meV. The Fermi level may be within those conically dispersing bands. 
%In that case, judging by the Seebeck coefficient measurement, the cones are at 100 meV away from the charge neutrality point.

%
%\section{Acknowledgments}
%
We acknowledge the help of D. LeBoeuf with high-field magnetoresistance measurements, and the use of high field facilities at LNCMI in Grenoble.
A.~A. acknowledges funding from the  Swiss National Science Foundation through project PP00P2\_170544.
Z.~R. was funded by the Postdoctoral Fellowship of the Swiss Confederation. Work of K.S. was supported by the Swiss National Science Foundation grant No. 200021\_175836.
M.N. and F.O. acknowledge the support of the Croatian Science Foundation under the project (IP-2018-01-8912) and CeNIKS project cofinanced by the Croatian Government and the EU through the European Regional Development Fund  - Competitiveness and Cohesion Operational Programme (Grant No. KK.01.1.1.02.0013).
M.N. acknowledges  the ISSP, University of Tokyo for partial financial support.
Work at Brookhaven National Laboratory was supported by the U.~S. Department of Energy, Office of
Basic Energy Sciences, Division of Materials Sciences and Engineering under Contract No.~DE-SC0012704.
%
% The bibliography (BibTeX)
%
\bibliography{TaIrTe4}

\end{document}